\documentclass[lengthcheck,twocolumn,prl,superscriptaddress,showpacs,floatfix,aps]{revtex4-2} 

\usepackage{fix-cm}
\usepackage{url}
\usepackage{amsmath,amssymb,amsfonts,latexsym,bm}
\usepackage{graphicx,epsfig}
\usepackage{fancyvrb}
\usepackage{booktabs}
\usepackage{soul}
\usepackage{algorithm}
\usepackage{algpseudocode}
\usepackage{ragged2e}
\usepackage[svgnames]{xcolor}
\usepackage[colorlinks,linkcolor=MediumBlue,citecolor=MediumBlue,urlcolor = MediumBlue]{hyperref}

\allowdisplaybreaks

\begin{document}

\fontsize{10.5pt}{12pt}\selectfont
\title{Policy heterogeneity improves collective olfactory search in 3-D turbulence} 

\author{Lorenzo Piro}\email{lorenzo.piro@roma2.infn.it}
\affiliation{Department of Physics \& INFN, Tor Vergata University of Rome, Via della Ricerca Scientifica 1, 00133 Rome, Italy}
\author{Robin A. Heinonen}
\affiliation{Department of Physics \& INFN, Tor Vergata University of Rome, Via della Ricerca Scientifica 1, 00133 Rome, Italy}
\affiliation{Machine Learning Genoa Center (MaLGa) \& Department of Civil,
Chemical and Environmental Engineering, University of Genoa, Genoa, Italy}
\author{Maurizio Carbone}
\affiliation{Istituto dei Sistemi Complessi, CNR, Via dei Taurini 19, 00185 Rome, Italy}
\affiliation{INFN ``Tor Vergata", Via della Ricerca Scientifica 1, 00133 Rome, Italy}
\author{Luca Biferale}
\affiliation{Department of Physics \& INFN, Tor Vergata University of Rome, Via della Ricerca Scientifica 1, 00133 Rome, Italy}
\author{Massimo Cencini}
\affiliation{Istituto dei Sistemi Complessi, CNR, Via dei Taurini 19, 00185 Rome, Italy}
\affiliation{INFN ``Tor Vergata", Via della Ricerca Scientifica 1, 00133 Rome, Italy}

\date{\today}

\begin{abstract}
     We investigate the role of policy heterogeneity in enhancing the olfactory search capabilities of cooperative agent swarms operating in complex, real-world turbulent environments. Using odor fields from direct numerical simulations of the Navier–Stokes equations, we demonstrate that heterogeneous groups, with exploratory and exploitative agents, consistently outperform homogeneous swarms where the exploration–exploitation trade-off is managed at the individual level. Our results reveal that policy diversity enables the group to reach the odor source more efficiently by mitigating the detrimental effects of spatial correlations in the signal. These findings provide new insights into collective search behavior in biological systems and offer promising strategies for the design of robust, bioinspired search algorithms in engineered systems.
\end{abstract}

\maketitle

\begin{figure*}[ht!]
\centering
\includegraphics[width=0.8\textwidth]{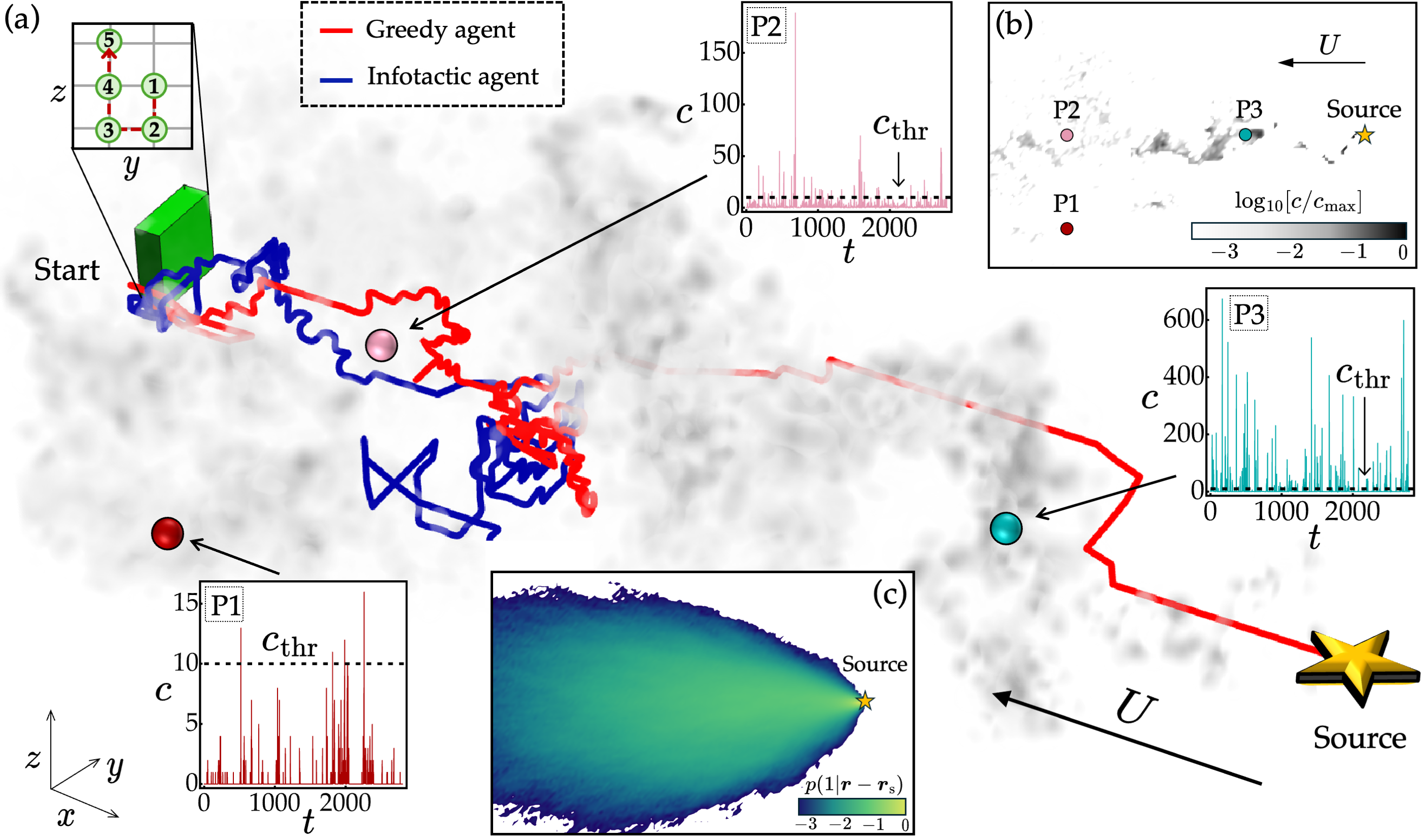}
\caption{\justifying (a) Snapshot of the concentration of odor particles (logarithmic grayscale) emitted by the source (yellow star), obtained by the direct numerical simulations of a turbulent flow as described in End Matter. The blue (red) curves correspond to the typical trajectories of an infotactic (greedy) agent starting from the region indicated by the green box and cooperating to find the source. Agents are initially placed close to each other, as depicted in the top-left inset. Three plots (P1-P3) display the intermittent time evolution of the odor signal at the point indicated by the arrow. In these plots, the dashed black line indicates the concentration threshold $c_{\rm thr}=10\approx2.4\bar{c}$  above which the agents can detect odor particles. Search starts when at least one agent detects an odor signal. Unless otherwise specified, hereafter, the mean wind strength $U$ with respect to the intensity of the turbulent fluctuations $u_{\rm rms}$ is equal to $U/u_{\rm rms}=2.4$.
(b) Two-dimensional snapshot of the concentration of odor particles within a thin slab parallel to the mean wind $U$ containing the source. Points P1, P2, and P3 are also reported here using the same color scheme as in the rest of the figure.
(c) Two-dimensional cut of the probability map of making a detection, $p(1|\bm{r}-\bm{r}_{\rm s})$, in the plane parallel to the mean wind $U$ and containing the source. This map, constituting the environment model the agents use to interpret their observations (empirical likelihood), is obtained by coarse-graining all the odors' trajectories, thresholding on the number of detectable particles ($c_{\rm thr}=10$), and finally averaging over time.}
\label{fig:1}
\end{figure*}

Looking for a target is a ubiquitous problem affecting ordinary life, from insects to humans. In the absence of any cue, the best searching agents can do is devise an efficient random search~\cite{benichou2011,tejedor2012,chechkin2018}.
In contrast, when cues are available, searchers need to optimally interpret them to find the target quickly. In particular, many biological and robotic agents face the challenge of locating an odor source in turbulent environments~\cite{murlis1992_review,wyatt2003,baker2018,francis2022_review}, where intermittent and patchy odor signals make strategies that work in simpler situations, such as gradient-climbing chemotaxis~\cite{berg1993random}, ineffective~\cite{crimaldi2001,balkovsky2002,celani2014} (see also Fig.~\ref{fig:1}).
Living organisms have evolved remarkable search strategies whose key to success is the balance between information gathering (exploration) and capitalizing on high-confidence cues (exploitation)~\cite{hernandezreyes2021,reddy2022_review}. This has inspired heuristic policies for individual agents, based on Bayesian inference~\cite{box2011}, that embody the exploration-exploitation trade-off, combining \emph{Infotaxis}~\cite{vergassola2007}, which seeks to maximize information gathering thus shifting the balance toward exploration, with policies biasing the search towards the likely source location~\cite{masson2013olfactory,loisy2022}. In particular, \emph{Space-Aware Infotaxis} (SAI)~\cite{loisy2022} refines this trade-off by blending Infotaxis with a \emph{Greedy} policy that aims to minimize the expected distance from the source~\cite{fernandez2006}. The robust performances achieved by SAI across diverse environments have established it as a state-of-the-art baseline for single-agent olfactory search~\cite{loisy2022,loisy2023,heinonen2023,heinonen2025}.

In nature, it has been widely observed that individuals significantly benefit from sharing information with their conspecifics~\cite{berdhal2018,nagy2020}, which inspired multi-agent extensions of heuristic search
strategies~\cite{karpas2017,song2019,durve2020,panizon2023}, so far mainly explored in simplified settings. However, understanding the exploration-exploitation trade-off for multi-agent systems in real-world turbulent environments is still an open theoretical problem, crucial to swarm robotics~\cite{francis2022_review,kwa2022} and applications ranging from environmental monitoring to rescue operations in harsh environments~\cite{verfuss2019,dang2020}.

Studies on animal behavior, foraging, and navigation have shown that groups of agents, ranging from small insects to large mammals, generally benefit from a division of labor, with some agents focusing on exploration, while others exploit the information collected by the group~\cite{seeley1982,holldobler1990,kengyel2015,goodale2024}.
In this Letter, inspired by these observations, we introduce policy heterogeneity as a way to balance exploration and exploitation in a group of cooperating agents searching for an odor source in a turbulent flow. Specifically, we show that heterogeneous swarms, combining infotactic and greedy agents, outperform homogeneous groups where all agents use SAI.
Our findings reveal that, under realistic environmental conditions, an optimal fraction of exploitative agents systematically speeds up source localization, while also reducing the probability of being lost, as compared to homogeneous swarms of SAI agents. 


{\it Collective olfactory search.} We consider a group of $N$ agents moving in a grid world $\Omega$ of size $L_x\!\times\! L_y\!\times\! L_z$ and lattice spacing $\Delta$. Their goal is to find a source, located in $\bm{r}_{\rm s}$, that sparsely emits odor particles whose trajectories are evolved with state-of-the-art DNS (see~\cite{piro2024} for details) of the incompressible Navier-Stokes equations under turbulent conditions ($\mathrm{Re}_\lambda \simeq 150$) with a uniform mean wind $\bm{U}$ along the $-\bm{\hat{x}}$ direction (Fig.~\ref{fig:1}).   The particle trajectories are coarse-grained within the 3-D arena, thus resulting in the concentration field $c$ displayed in grayscale in Fig.~\ref{fig:1}. 
Agents have a prescribed sensitivity to the odor and make binary observations, i.e., a detection occurs only when the highly fluctuating and intermittent local concentration (see panels P1-P3 in Fig. \ref{fig:1}) exceeds a given threshold $c_{\rm thr}$.

We ensure that each search episode starts with at least one agent detecting the odor signal. Then, the remaining agents are placed in a neighborhood of the first (see top-left inset in Fig.~\ref{fig:1}(a) and End Matter for details on the numerical setup). All agents share a probability spatial map (a ``belief'') about the position of the source defined over $\Omega$, $b(\bm{r})\equiv \mathrm{Prob}(\bm{r}_{\rm s}=\bm{r})$. Assuming no prior knowledge, the belief is initialized to a uniform distribution and set to zero at the agents' positions. Within the Bayesian framework, agents must rely on a model of the environment to interpret their measurements and localize the source, reflecting the idea that the agents possess prior knowledge of the odor signal's statistical characteristics. We therefore assume that each agent $i$ knows its position, $\bm{r}_i$, and the {\it likelihood} 
$p(h_i^{(t)}|\bm{r}_i-\bm{r})$ of making an observation, $h_i^{(t)}\in\{0,1\}$, at time $t$ obtained from the time average of the coarse-grained DNS data. This model of the environment (see Fig.~\ref{fig:1}(c) for a 2-D projection) only captures the single-time odor statistics, as it neglects the spatiotemporal correlations exhibited by the odor plume, and is therefore not ``exact.'' However, it represents the best tractable model of the turbulent odor signal generated from the DNS and will hereafter be referred to as the \emph{empirical likelihood}.

At each time step, the agents synchronously measure the odor concentration at their positions and update the shared belief via Bayes’ rule~\cite{box2011}
\begin{equation}
    b^{(t)}(\bm{r})=\frac{1}{\mathcal{N}}\prod\limits_{i=1}^{N}p(h_i^{(t)}|\bm{r}_i-\bm{r})b^{(t-1)}(\bm{r}) \, ,
    \label{eq:bayes}
\end{equation}
where $\mathcal{N}$ is a normalization factor ensuring the belief integrates to one over the domain $\Omega$. Notice that to have a shared belief, agents only need to communicate their positions and detections ($0/1$), feasible in swarm robotics by wireless communication~\cite{girma2020iot}. 
The cost of communication is thus greatly reduced with respect to communicating the entire belief. Upon updating it via~\eqref{eq:bayes}, at each time step, each agent chooses an action, that is, to move in one of the allowed directions \{$\pm\bm{\hat{x}},\pm\bm{\hat{y}},\pm\bm{\hat{z}}$\} by a lattice spacing $\Delta$.

Given the complexity of multi-agent olfactory search, it is practically impossible to identify an optimal collective protocol.
Thus, in the following, we focus on comparing heuristics that embody the exploration-exploitation trade-off of the group of agents in two different ways.
In particular, inspired by the division of labor \cite{kengyel2015}, we consider
a swarm of agents divided into two groups; Group (A) follows a risk-adverse \emph{Infotactic} strategy~\cite{vergassola2007}, which prioritizes exploration by selecting actions that maximize the expected information gain about the source location; Group (B) follows a strongly exploitative \emph{Greedy} policy~\cite{fernandez2006}, meant to reduce the expected Manhattan distance to the source. Although both strategies perform a one-step planning based on the current belief, which summarizes the information gathered until that time, for greedy agents, the absence of an exploratory component often results in suboptimal trajectories, such as getting trapped in local maxima far from the actual source~\cite{loisy2022}.
We shall compare the performance of our heterogeneous group of $N$ agents made of different fractions of $N_A$  and $N_B=N-N_A$ players against the baseline where all agents manage the exploration–exploitation trade-off at the individual level, considering a homogeneous swarm where each agent adopts the \emph{Space-Aware-Infotaxis} (SAI) policy~\cite{loisy2022,panizon2023}, which combines Infotaxis and Greedy policy as described below.

Depending on its state $s_i\!\equiv\!(\bm{r}_i,b(\bm{r}))$, defined by its position and the shared belief, and on its own policy, $\pi_i$, the $i$th agent selects the action that minimizes a policy-dependent cost function $\mathcal{C}_{\pi_i}$:
\begin{equation}
    \label{eq:policies}
    a^*_i={\rm argmin}_a\mathcal{C}_{\pi_i}(s_i,a) \; , \; \; \rm with
\end{equation}
\begin{equation}
    \label{eq:costs}
    \mathcal{C}_{\pi_i}(s_i,a)= \begin{cases}
                        \frac{1}{2}(2^{H(s_i|a)}-1) \; , \hspace{0.65cm} \rm if \; \pi_i=\rm Infotaxis\\
                        D(s_i|a) \; , \hspace{1.85cm} \rm if \; \pi_i=\rm Greedy \\
                        \mathcal{C}_{\rm Infotaxis}+\mathcal{C}_{\rm Greedy} \; ,\; \rm if \; \pi_i=\rm SAI \, ,
                    \end{cases}
\end{equation}
where $H(s_i)\!\equiv\!-\!\sum_{\bm{r}}b(\bm{r})\log_2 b(\bm{r})$ is the Shannon entropy of the shared belief, and $D(s_i)\!\equiv\! \sum_{\bm{r}}b(\bm{r})||\bm{r}_i-\bm{r}||_1$ the expected Manhattan distance from the source (see also~\cite{vergassola2007,loisy2022,panizon2023,heinonen2023,heinonen2025}). All agents move synchronously, measure the odor concentration, and update the shared belief again, repeating until one of them satisfies $\bm{r}_i=\bm{r}_{\rm s}$, indicating that the source has been found. If the search exceeds a predefined time limit ($T_{\rm max}$), the episode ends with the agents considered lost.
Although actions are not coordinated among agents~\cite{masson2009,karpas2017}, as it would increase computational complexity exponentially with the number of agents, cooperation is ensured by sharing the belief~\cite{masson2009,panizon2023}. In addition, an action is forbidden if it brings the agent outside the domain $\Omega$ or makes it overlap with another agent; in the absence of allowed actions, the agent stands still. See End Matter for more details on the numerical implementation of the collective search algorithm and on the heuristic policies introduced above.

\begin{figure}[t!]
\centering
\includegraphics[width=\columnwidth]{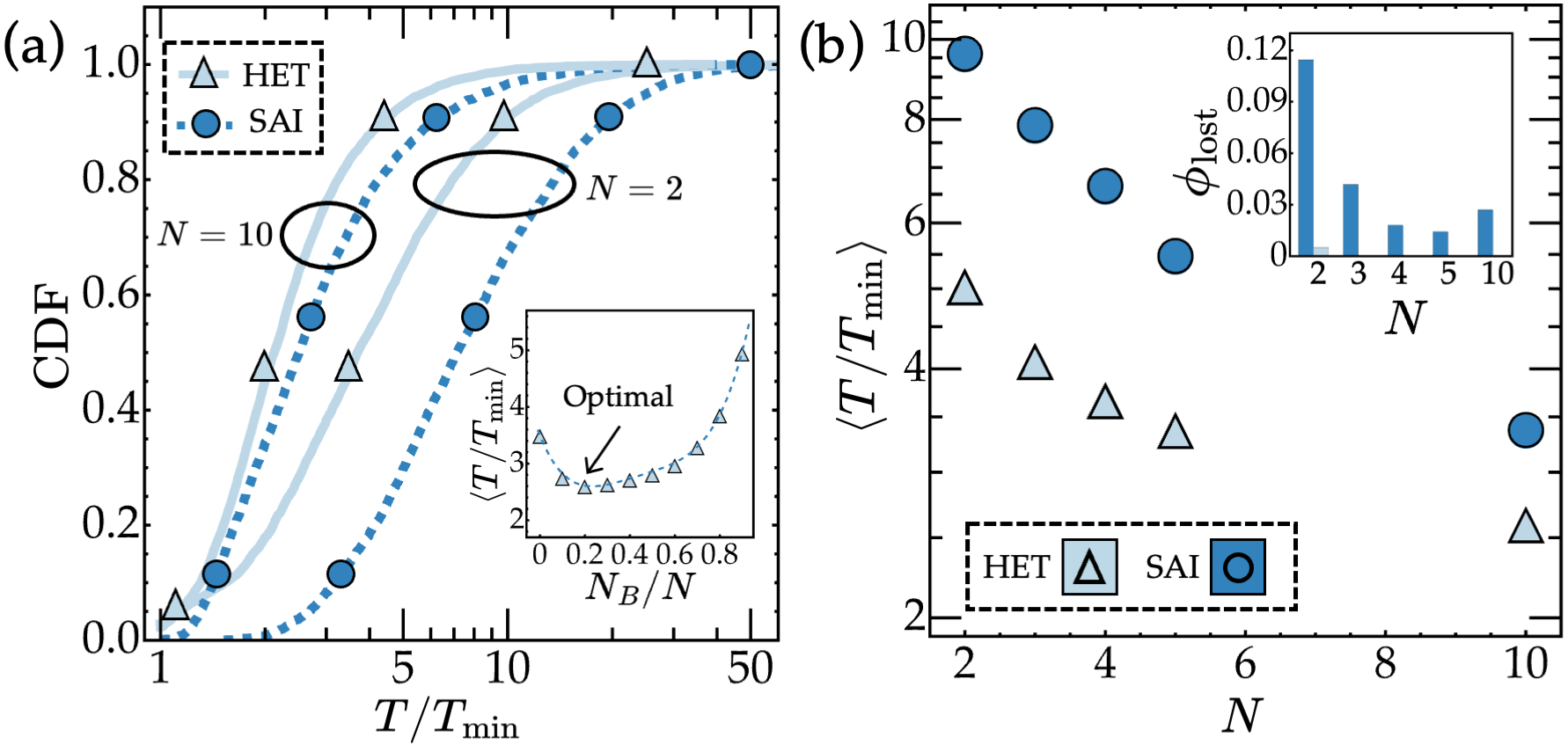}
\caption{\justifying (a) Cumulative distribution function (CDF) of first agent arrival times $T$ normalized with the minimum search time $T_{\rm min}$ (agent's initial Manhattan distance from the source) when using two or ten SAI agents (dashed dark blue curves) or heterogeneous (HET) swarms with the same number of agents (solid light blue curves). The results of the HET swarms correspond to those obtained with the optimal fraction of greedy agents $N_B/N$, i.e., the configuration minimizing the mean arrival time to the source, as shown in the inset for $N=10$ agents. (b) Mean first arrival times as a function of the number of agents $N$ composing the SAI swarm (dark blue circles) or the optimal HET configuration (light blue triangles). The arrival time statistics shown here are conditioned on successful episodes, while the inset shows the fraction of episodes $\phi_{\rm lost}$ in which the search fails ($T>T_{\rm max}=2\cdot10^3$). Note that only the lost fraction for SAI is visible since for HET swarms $\phi_{\rm lost}<0.01$. Data presented here have been obtained by averaging over $5250$ episodes.}
\label{fig:2}
\end{figure}

{\it The benefit of policy heterogeneity.}
Figure~\ref{fig:2}(a) shows the cumulative distribution functions (CDF) of the first agent arrival time to the source conditioned on successful episodes, i.e., those in which agents do not get lost ($T<T_{\rm max}\!=\!2\cdot10^3$), for HET and SAI swarms made up of $N\!=\!2$ and $N\!=\!10$ agents, respectively. For each $N$, we find that an optimal fraction of greedy agents exists in HET swarms that maximizes the overall search efficiency (see inset for $N\!=\!10$), making them achieve shorter first arrival times to the source compared to swarms comprising only SAI agents. 
Indeed, as demonstrated in Fig.~\ref{fig:2}(b), HET swarms consistently show $\gtrsim 25\%$ improvement in terms of mean first arrival time with respect to SAI agents and, remarkably, the performance achieved with 10 SAI agents can be achieved with a HET group of just 5 agents ($N_A\!=\!3$ and $N_B\!=\!2$). Moreover, the latter always manages to reach the source in a finite time, whereas SAI swarms have a finite probability of getting lost during the search (inset of Fig.~\ref{fig:2}(b)). Consistently, the greedy agents in a HET group are the most likely to find the source. For instance, in the optimal HET configuration with $N\!=\!10$ agents ($N_A\!=\!8$ and $N_B\!=\!2$), despite being only two out of ten in the group, the greedy agents reach the source first in $63\%$ of the episodes.

\begin{figure}[ht!]
\centering
\includegraphics[width=\columnwidth]{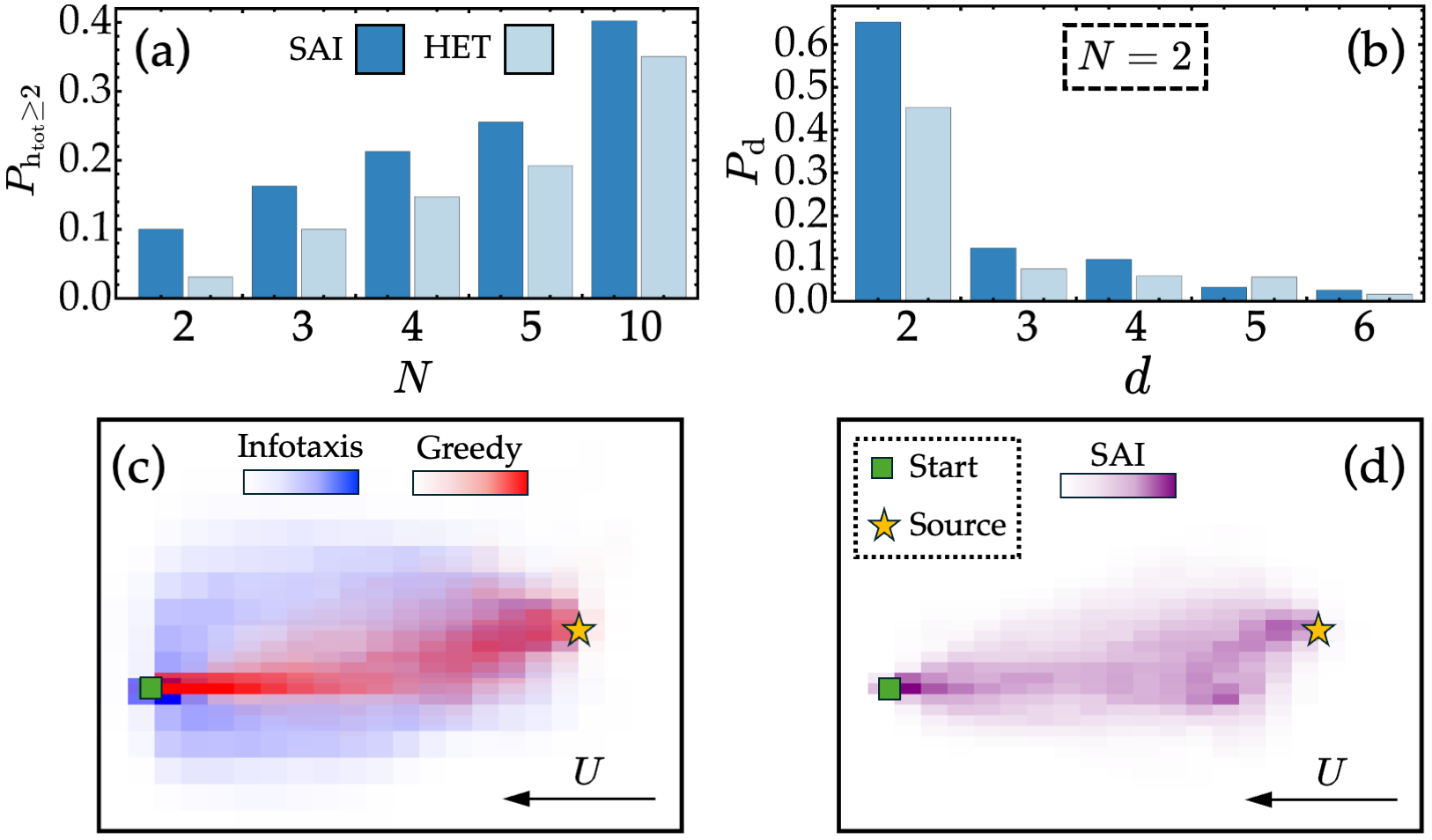}
\caption{\justifying (a) Probability, $P_{\rm h_{tot}\geq2}$, that, at any time during the search, two or more agents simultaneously detect an odor signal versus the swarm size $N$. (b) Probability, $P_{\rm d}$, of the distance $d$ between two agents when these make a simultaneous detection. In both (a) and (b), the probabilities are conditioned on successful episodes. (c)-(d) Heatmaps of the trajectories of ten agents starting from the region indicated by the green square that find the source (yellow star) when $N_A=N_B=N/2$ (c) or when they are all using SAI (d). The arrow indicates the mean wind direction. Such maps have been obtained in the 2-D setup with agents' detection threshold $c_{\rm thr}=10$ and $U/u_{\rm rms}=2.4$ (in units of the intensity of the turbulent fluctuations), and averaging over 440 episodes. Color gradient indicates the paths' density.}
\label{fig:3}
\end{figure}

{\it Role of turbulent correlations.} 
A key factor behind the observed improvement of HET swarms over SAI is how the two groups manage the presence of spatiotemporal correlations in the odor signal. In realistic turbulent flows, odor plumes tend to form filaments and puffs (Fig.~\ref{fig:1}), potentially leading to highly correlated detections between nearby agents. These events are not included in the model likelihood (Fig.~\ref{fig:1}(c)) the agents use to interpret their observations, as it only accounts for the single-point statistics.
To quantify the role of correlations, we measure the probability of making simultaneous detections across SAI and HET swarms. Figure~\ref{fig:3}(a) shows that SAI agents systematically experience a higher frequency of concurrent detections compared to their HET counterparts, with such events mostly happening when the agents are just a few lattice points away from each other, as reported in Fig.~\ref{fig:3}(b). 
Moreover, the two-dimensional (2-D) heatmaps shown in Fig.~\ref{fig:3} qualitatively demonstrate how HET swarms, thanks to their diversity in the action selection, tend to spread out more, better covering the space than SAI agents (compare Fig.~\ref{fig:3}(c) and (d) and see movies in Supplemental Material (SM)~\footnote{See {S}upplemental {M}aterial [url], for supplementary movies and figures.} showcasing typical 3-D agents' trajectories, thus providing visual evidence of such behavioral differences between HET and SAI swarms). 
This analysis reveals that agents in a homogeneous SAI group are more prone to clustering, resulting in more likely redundant detections, due to correlations, and consequently inefficient exploration. Conversely, a mix of greedy and infotactic agents promotes greater movement diversity, preventing unnecessary clustering, thereby enhancing the overall efficiency of the search. 
In Fig.~S1 of SM~\cite{Note1}, we further support these findings through the statistical analysis of blanks, i.e., the time intervals between two consecutive odor detections in a swarm, emphasizing the impact of odor intermittency on search efficiency. There, we also highlight that when odor detections are generated directly from the same empirical likelihood used by the agents to interpret them, the performance gap between SAI and HET swarms, as well as the observed disparity in the correlated detections and average duration of the blanks, significantly narrows and tends to vanish as group size increases. This emphasizes the benefit of policy heterogeneity when agents observe a realistic turbulent odor signal and further underscores the role of correlations and odor intermittency in shaping collective olfactory search dynamics. It is also consistent with the already observed quasi-optimality of SAI in both individual~\cite{loisy2022} and collective~\cite{panizon2023} olfactory search in simplified settings.

\begin{figure}[ht!]
\centering
\includegraphics[width=\columnwidth]{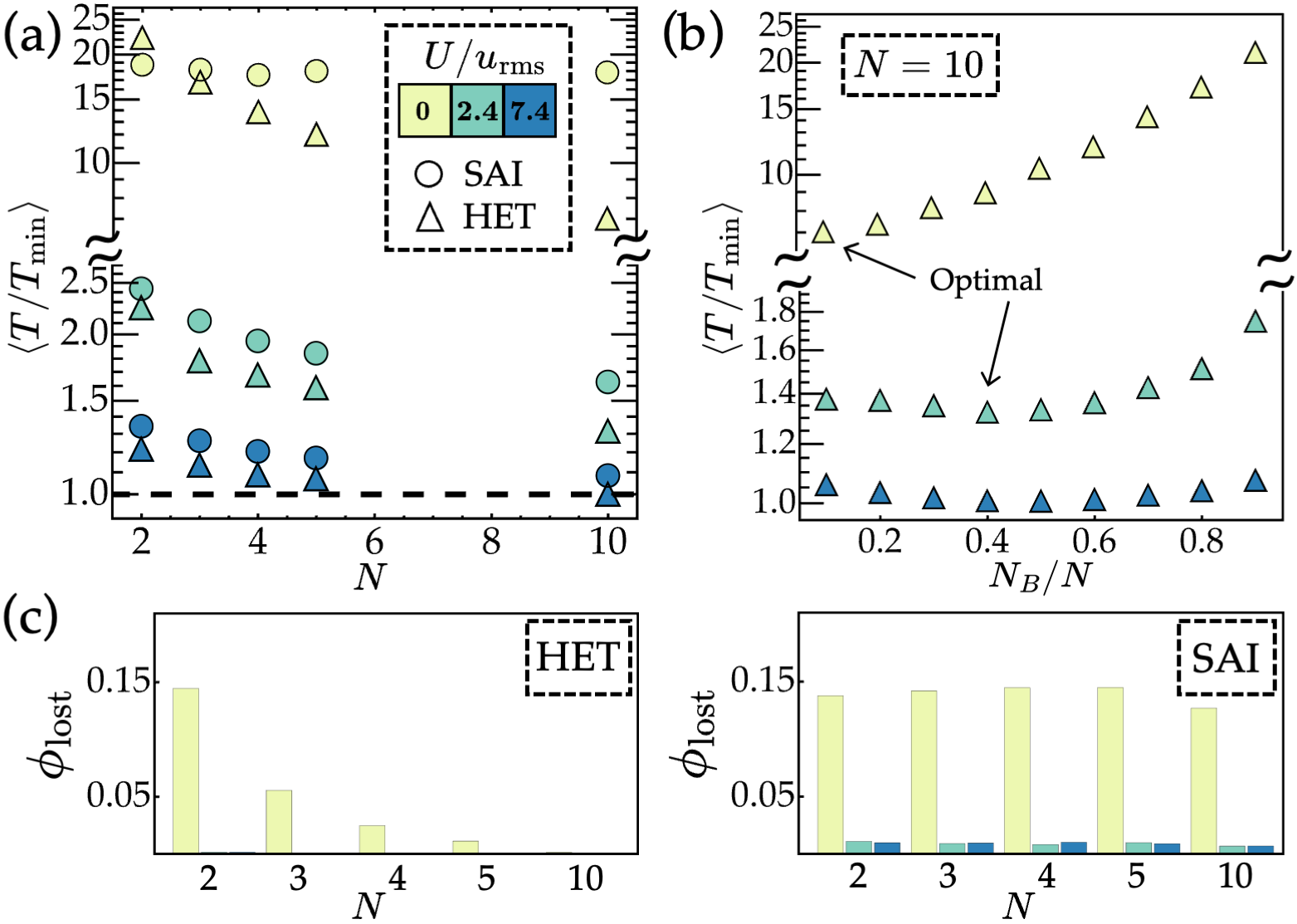}
\caption{\justifying (a) Normalized mean arrival times
$\langle T/T_{\rm min}\rangle$ ($T_{min}$ being the initial Manatthan distance from the source) conditioned on successful episodes as a function of the number of agents $N$ composing the SAI swarm (circles) or the optimal HET configuration (triangles). Colors distinguish the strength of the mean wind $U$ with respect to the $u_{\rm rms}$ of the turbulent flow as in legend.
The dashed black line denotes the theoretical optimum $\langle T/T_{\rm min}\rangle=1$. (b) Mean normalized arrival time as a function of the fraction of greedy agents $N_B/N$ in heterogeneous swarms for $N=10$. (c) Fraction of episodes $\phi_{\rm lost}$ in which agents get lost (i.e. $T>T_{\rm max}=1400$) for both HET (left) and SAI swarms (right). Color legends in (b)-(c) are the same as (a). Data have been averaged over $10^4$ episodes.}
\label{fig:4}
\end{figure}

{\it Comparison across different environments.} 
To scrutinize the robustness of our findings against environmental conditions, we investigate how different wind conditions influence the effectiveness of HET and SAI swarms. We do this by restricting the search to 2-D, thus allowing the agents to move within a thin slab containing the source and aligned with the mean wind direction, when present. 
The results obtained under three different wind intensities are summarized in Fig.~\ref{fig:4} (Fig.~S2 in SM~\cite{Note1} shows the empirical likelihoods in the three scenarios).

We find that increasing the wind strength simplifies the search task by providing agents with stronger directional cues about the source location and restricting the detection cone. Although this tends to diminish the advantage of policy heterogeneity, HET swarms still systematically outperform SAI ones, managing to find the source in the shortest possible time $T_{\rm min}$ (dashed black line in Fig.~\ref{fig:4}(a)). In the absence of a mean wind, correlations are stronger and the information provided by the model is lower, making the search task more challenging. In this scenario, the performance gap between HET and SAI swarms further expands. In fact, deploying more SAI agents reduces neither the mean arrival time (Fig.~\ref{fig:4}(a)) nor the probability of getting lost (Fig.~\ref{fig:4}(c)), which remains constant at $\simeq 14\%$. In contrast, the mean arrival time of HET swarms improves by a factor of three when passing from two to ten agents, with a vanishing lost fraction $\phi_{\rm lost}$. This further emphasizes the benefit of policy heterogeneity in overcoming the challenges posed by turbulent odor correlations. In addition, these results depend only weakly on the sensitivity of the agents to the odor signal and remain qualitatively unchanged when varying the detection threshold $c_{\rm thr}$ (Fig.~S3 in SM~\cite{Note1}).

Remarkably, we find that the optimal fraction of greedy agents depends on the complexity of the task. As shown in Fig.~\ref{fig:4}(b), more challenging (i.e., less informative) search scenarios (weaker winds) require fewer greedy agents to maintain an effective balance between exploration and exploitation. Still, greedy agents prevalently find the source, e.g., for $N=10$ and no mean wind, the single greedy agent reaches the source in $28\%$ of the episodes despite representing $10\%$ of the swarm.

{\it Discussion.} 
Our study reveals that introducing policy heterogeneity within a swarm of cooperating agents significantly improves the efficiency of collective olfactory search in realistic turbulent environments, both in terms of search time and robustness. As a key insight, we find that heterogeneity mitigates the challenges posed by turbulent intermittency and correlations (not included in the model of the environment used by the agents) of the odor signal, enabling swarms to balance exploration and exploitation more effectively than collectively using the same strategy. 
Significantly, heterogeneity benefits both the individual and the entire group by optimizing the mean arrival time of both the first and last agent (see Fig.~S4 in SM~\cite{Note1}), thus enhancing the overall efficiency and coordination of the swarm and ensuring that all agents reach the source more quickly.

Our results, robust in 2-D/3-D settings with different environmental conditions and detection sensitivity, align with biological observations of various search behaviors displayed by animals, suggesting that policy heterogeneity might be a natural optimization mechanism~\cite{traniello1997,duarte2012,jolles2020,frolich2025}. Likewise, these insights could inform the design of artificial swarm systems for real-world applications such as environmental monitoring or disaster response~\cite{dorigo2013,khaldi2025}.

Based on the insights provided by our work, future research could explore adaptive learning mechanisms, allowing agents to develop mixed strategies~\cite{barraclough2004prefrontal} to dynamically switch between exploratory and exploitative policies in response to real-time environmental feedback. Furthermore, given the key role played by spatiotemporal correlations present in the olfactory signal, their incorporation into the environment model could provide valuable insight into optimizing heterogeneous swarms in realistic turbulent settings (see~\cite{heinonen2025} for recent efforts in this direction). 

Finally, demonstrating the advantage of policy heterogeneity, our results lay the foundations for further improvements in collective olfactory search. Indeed, our approach opens new avenues for mixing additional strategies to achieve an even sharper exploration–exploitation balance. Moreover, future work could integrate heterogeneity into model-free methods~\cite{singh2023,rando2025}, allowing for refining and adapting the swarm behavior dynamically via Multi-Agent Reinforcement Learning~\cite{albrecht2024}; our results establish a strong baseline for comparison.


\acknowledgments
We thank Antonio Celani and Massimo Vergassola for useful discussions and their insightful comments on the first draft of this Letter. We acknowledge financial support under the National Recovery and Resilience Plan (NRRP), Mission 4, Component 2, Investment 1.1, Call for tender No. 104 published on 2.2.2022 by the Italian Ministry of University and Research (MUR), funded by the European Union --- NextGenerationEU --- Project Title Equations informed and data-driven approaches for collective optimal search in complex flows (CO-SEARCH), Contract 202249Z89M. --- CUP B53-D23003920006 and E53-D23001610006. This work was also supported by the European Research Council (ERC) under the European Union’s Horizon 2020 research and innovation program (Grant Agreement Nos.\ 882340 and 101002724), by the Air Force Office of Scientific Research (grant FA8655-20-1-7028), and the National Institute of Health under grant R01DC018789. 

\appendix

\setcounter{figure}{0}
\renewcommand\thefigure{A\arabic{figure}}
\setcounter{equation}{0}
\renewcommand\theequation{A\arabic{equation}}

\section{END MATTER} 

\subsection{Details on the numerical simulations} 
\label{sec:1}

In our numerical simulations, we coarse-grained the concentration of odor particles obtained from the DNS (detailed in \cite{piro2024,heinonen2025}) within a 3-D volume $\Omega$ of size $129\Delta\times99\Delta\times99\Delta$, with $\Delta\approx12\eta$ being the lattice spacing and $\eta$ the Kolmogorov scale. Time is always expressed in units of observation time of the agents, which is here set equal to the Kolmogorov timescale $\tau_\eta$ of the turbulent flow. The source is placed in $\bm{r}_{\rm s}=(115,49,49)$, with the mean wind blowing in the $-\hat{\bm{x}}$ direction. At the beginning of each episode, the initial position of the first agent is drawn uniformly from any of the points $\bm{r}\in\Omega$ where the concentration exceeded the detection threshold $c_{\rm thr}=10\approx2.4\bar{c}$ at a randomly selected time $t_s$ within the DNS. All the data presented in the 3-D setup have then been obtained by averaging the results over $5250$ episodes.

In the 2-D setup, we coarse-grained the concentration of particles within a thin slab of thickness $\Delta\approx12\eta$ containing the source. The resulting 2-D arena has an extension of $129\Delta\times129\Delta$. When in the presence of a mean wind $\bm{U}$ (oriented in the $-\hat{\bm{x}}$ direction), the source is placed in $\bm{r}_{\rm s}=(115,64)$, otherwise, in the isotropic case ($U=0$), $\bm{r}_{\rm s}=(64,64)$. The results obtained in the 2-D setup are averaged over $10^4$ episodes.

A detailed step-by-step description of the collective olfactory search algorithmic procedure is given in Algorithm~\ref{alg:multiagent}.

\begin{algorithm}[H]
\caption{Collective olfactory search with shared belief and synchronous update}
\label{alg:multiagent}
\begin{algorithmic}[1]
\For{$j=1$ to $N_{\rm epis}$} \Comment{Loop over episodes.}
    \State Select random time $t_s$ in the DNS.
    \State Select random position $\bm{r}_0\in\Omega$ s.t. $c(\bm{r}_0,t_s)>c_{\rm thr}$.
    \State Place first agent in $\bm{r}_0$.
    \State Arrange remaining agents around the first.
    \State Initialize shared belief $b^{(0)}$ to a uniform distribution. \State Set $b^{(0)}$ to zero in the agents' positions and renormalize.
    \While{found=false and $t<T_{\rm max}$} \Comment{Time loop.}
        \State Agents measure $\bm{h}^{(t)}$.
        \State Update belief via Bayes' rule [Eq.(1)].
        \State Agents pick $a^*$ based on state and policy [Eqs.(2-3)].
        \State Agents move.
        \If {$\bm{r}_i=\bm{r}_s$}           \Comment{Check if found source.}
            \State found=true                    
            \State $T_{\rm found}^{(j)}=t$        \Comment{Output first arrival time of each episode.}
        \EndIf
    \State $t\to t+1$
    \EndWhile
\EndFor
\end{algorithmic}
\end{algorithm}

\subsection{Details on the heuristic policies}
\label{sec:2}


In the main text, we have analyzed three heuristic policies that determine the agents’ decision-making based on their model of the environment and the information accumulated in the shared belief, via their binary detections of the odor plume. 

All three policies rely on a simple yet effective operational rule: at each step, based on its current position and the shared belief, each agent chooses (independently, yet synchronously) the action among those allowed (as mentioned in the main body we forbid exit from $\Omega$ and co-occupation of the same position by two or more agents) that minimizes a cost function that depends on the assigned policy.

We refer the reader to Refs.~\cite{fernandez2006,vergassola2007,masson2013olfactory,loisy2022,panizon2023} for further details on the heuristic strategies illustrated below.

\paragraph*{\textbf{Infotaxis~\cite{vergassola2007} ---}}

The key principle behind Infotaxis is to maximize information gathering, i.e., the expected reduction in entropy $H$ of the belief $b(\bm{r})$, which represents the probability distribution of the source location over the search space. Here, the entropy thus quantifies the uncertainty in the target's position and is defined as
\begin{equation}
    H(s_i)\equiv-\sum_{\bm{r}}b(\bm{r})\log_2 b(\bm{r}) \, ,
    \label{eq:S1}
\end{equation}
where $s_i\equiv(\bm{r}_i,b(\bm{r}))$ is the state of the $i-$th agent. At each step, the $i-$th infotactic agent selects the action $a^*_i$ that is expected to yield the highest information gain, quantified as the decrease in entropy of the belief distribution, i.e., the action that minimizes the cost
\begin{equation}
    \mathcal{C}_I(s_i,a) = H(s_i|a) - H(s_i) \, .
    \label{eq:S2}
\end{equation}
Note that minimizing this cost is indeed equivalent to maximizing the mutual information between the action $a$ and the current state $s_i$.

Importantly, as originally demonstrated by Vergassola et al.~\cite{vergassola2007}, the entropy of the belief is correlated with the search time. Indeed, it is possible to show that, given a belief with entropy $H$, a lower bound for the time to find the source would be $\sim \exp(H-1)$. We may thus rewrite the infotactic cost so as to highlight such a connection (see also~\cite{loisy2022}):
\begin{equation}
    \mathcal{C}_{\rm I}(s_i,a) = \frac{1}{2}(2^{H(s_i|a)} - 1) \, ,
    \label{eq:S3}
\end{equation}
This is a monotonous function that preserves the ordering in $a$, such that actions that minimize the cost~\eqref{eq:S2} will also correctly minimize~\eqref{eq:S3}. Furthermore, this formulation underscores how lower entropy correlates with reduced exploration time in idealized scenarios.

Following this strategy, the agent will naturally shift, in the course of time, from exploring areas of high uncertainty to homing in on the likely source location once sufficient information has been gathered.\\

\paragraph*{\textbf{Greedy~\cite{fernandez2006} ---}}

The greedy search policy follows a more straightforward approach by always moving toward the direction that minimizes the expected Manhattan distance from the source, which reads
\begin{equation}
    D(s_i) = \sum_{\bm{r}}b(\bm{r})||\bm{r}_i-\bm{r}||_1 \, .
    \label{eq:S4}
\end{equation}
Therefore, the cost function that the $i-$th greedy agent aims to minimize takes the simple form:
\begin{equation}
    \mathcal{C}_{\rm G}(s_i,a)= D(s_i|a) \, .
    \label{eq:S5}
\end{equation}
This purely exploitative strategy can be effective in environments where odor cues are strong and reliable. However, greedy search is prone to failure in turbulent environments as it does not account for the uncertainty or potentially misleading nature of odor detections. It may thus lead the agent into regions not necessarily close to the actual source, resulting in inefficient or stalled searches. This already happens when the agent deploys the correct model of the environment~\cite{loisy2022}, and it gets even more severe when the model is inexact, as in our case.\\

\paragraph*{\textbf{Space-Aware Infotaxis~\cite{loisy2022} ---}}

Space-aware Infotaxis (SAI) extends the standard Infotaxis framework by incorporating spatial awareness into the decision-making process. While Infotaxis optimizes for informational gain alone, SAI introduces an additional term that accounts for the expected Manhattan distance to the target. In other words, the cost function that the $i-$th SAI agent aims at minimizing is the linear combination of the infotactic~\eqref{eq:S3} and greedy~\eqref{eq:S5} one, i.e.,
\begin{equation}
    \mathcal{C}_{\rm SAI}(s_i,a) = \frac{1}{2}(2^{H(s_i|a)} - 1) + D(s_i|a) \, .
    \label{eq:S6}
\end{equation}
The guiding principle is that when uncertainty about the source location is high, prioritizing information gain is beneficial. However, as confidence in the source location increases, minimizing the distance to the source becomes more critical. The specific mathematical formulation of the SAI cost function reflects this trade-off, ensuring that the agent transitions smoothly from exploration to exploitation, improving search efficiency of the single agent without excessive detours.\\

\clearpage

\section{Supplementary movies and figures}

\textbf{Caption of \emph{movie$\_$3D$\_$HETvsSAI.mp4}:} The movie shows the comparison between the trajectories of $N=10$ SAI agents (right panel) and $N=10$ HET agents with the optimal fraction of greedy agents (left panel), i.e., comprising eight infotactic agents (blue) and two greedy (red). Both swarms start at the same time and from the same region indicated by the green box and cooperate to find the odor source, here denoted with a red sphere. The concentration of odor particles (shown in logarithmic grayscale) emitted by the source is the one obtained by direct numerical simulations of the 3-D Navier-Stokes equations.
Despite initially observing the same signal, the two groups take different paths according to their own policy. 
In fact, this visualization highlights key behavioral differences: greedy agents (left panel, red) tend to move directly upwind, immediately exploiting the information gathered by the group, while infotactic agents (left panel, blue) are more cautious and therefore explore more extensively the space moving crosswind. This partitioning of labor helps the whole group achieve a better and more robust performance, as they reach the source systematically faster than homogeneous SAI swarms. On the other hand, although SAI agents also start by exploring the environment, they then tend to cluster, significantly slowing down the search as they become increasingly influenced by correlations in the odor signal.
Note that consecutive frames are spaced apart by twice the duration of the agents' observation, which is equal to the Kolmogorov timescale $\tau_\eta$ of the flow. Also, to avoid clutter, we display only the trail corresponding to the last ten time steps for each agent.\\

\textbf{Caption of \emph{movie$\_$3D$\_$HETvsSAI$\_$hardEpisode.mp4}:} Same as \emph{movie$\_$3D$\_$HETvsSAI.mp4}, but in an episode where the SAI swarm gets lost, while the HET group manages to find the source in a finite time. Legends are the same as in \emph{movie$\_$3D$\_$HETvsSAI.mp4}.

\begin{figure*}[h!]
\centering
\includegraphics[width=0.8\textwidth]{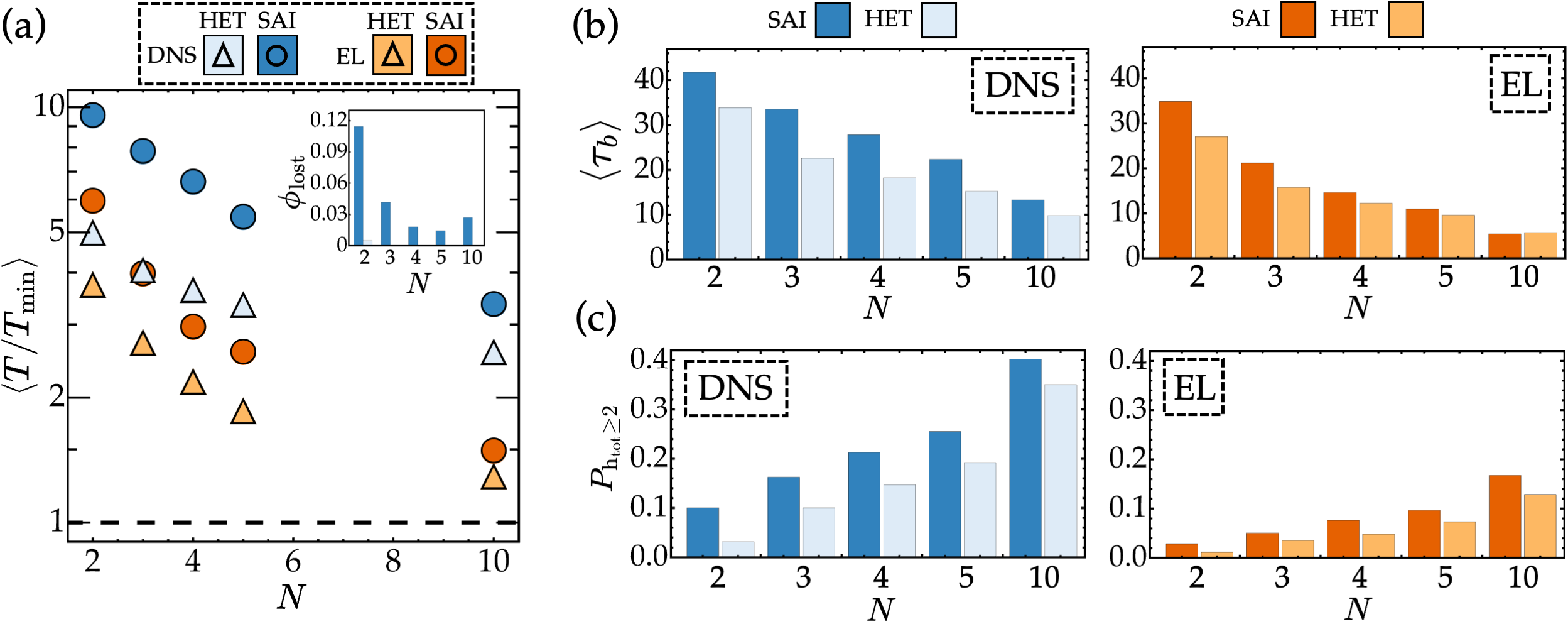}
\caption{\justifying \textbf{Comparison with the exact model.} (a) Mean arrival times in units of the minimum search time $T_{\rm min}$ (agent's initial Manhattan distance) as a function of the number of agents $N$ composing the SAI swarm (circles) or the optimal heterogeneous policy configuration (triangles). Blue shades indicate the results obtained using the DNS signal, while orange shades denote the outcome achieved when deploying the exact model of the environment, i.e., when the empirical likelihood (EL) is used to generate odor encounters. That is, at every time step $t$, each $i-$th agent makes a detection $h^{(t)}_i=\{0,1\}$ based on a binomial process with success probability equal to the EL value at its current position.
The dashed black line corresponds to the theoretical optimum $\langle T/T_{\rm min}\rangle=1$. The fraction of episodes $\phi_{\rm lost}$ failing to find the source in the allotted time $T_{max}=2\cdot 10^3$ is reported in the inset. Note that only SAI in the DNS setup (dark blue bars) is visible since all other cases feature $\phi_{\rm lost}<0.01$.  
As it turns out, the performance gap between SAI and HET swarms significantly narrows when using the exact model and tends to vanish as group size increases.
(b) Average duration of blanks $\langle \tau_b\rangle$, i.e., the time intervals between two consecutive odor detections in a swarm, in units of measurements made by the agents, and (c) probability $P_{\rm h_{tot}\geq2}$ that, at any time during the search, two or more agents make a detection simultaneously as a function of the swarm size $N$. The color legend is the same as in (a).
Some observations are in order. First, SAI swarms feature both a larger $\langle \tau_b\rangle$ and $P_{\rm h_{tot}\geq2}$ than HET ones when observing the realistic signal obtained from the DNS, regardless of the number of agents (compare dark blue with light blue histograms). This is a double indication that SAI groups are more strongly affected by the sparseness and spatiotemporal correlations of the odor signal, which is indeed reflected in their poorer performance in the source localization task. On the other hand, when using the exact model ---thus drawing the detections from the empirical likelihood (EL)--- such differences between SAI and HET swarms tend to thin out (compare dark orange with light orange histograms), which is consistent with the observed narrowing of the arrival time performance gap [see orange symbols in panel (a)]. In particular, the probability of making simultaneous detections, $P_{\rm h_{tot}\geq2}$, becomes significantly smaller, consistent with the fact that the odor signal is uncorrelated in this case.
This further highlights the benefit of policy heterogeneity when agents observe the actual turbulent odor signal produced by the DNS, underscoring the importance of spatiotemporal correlations in designing robust collective search strategies.
Note that all the data shown here are conditioned on successful episodes in which the agents do not get lost, i.e., $T<T_{\rm max}=2\cdot10^3$, and have been averaged over 5250 episodes.
}
\label{fig:dnsVSempLk_noLost}
\end{figure*}

\begin{figure*}[ht!]
\centering
\includegraphics[width=\textwidth]{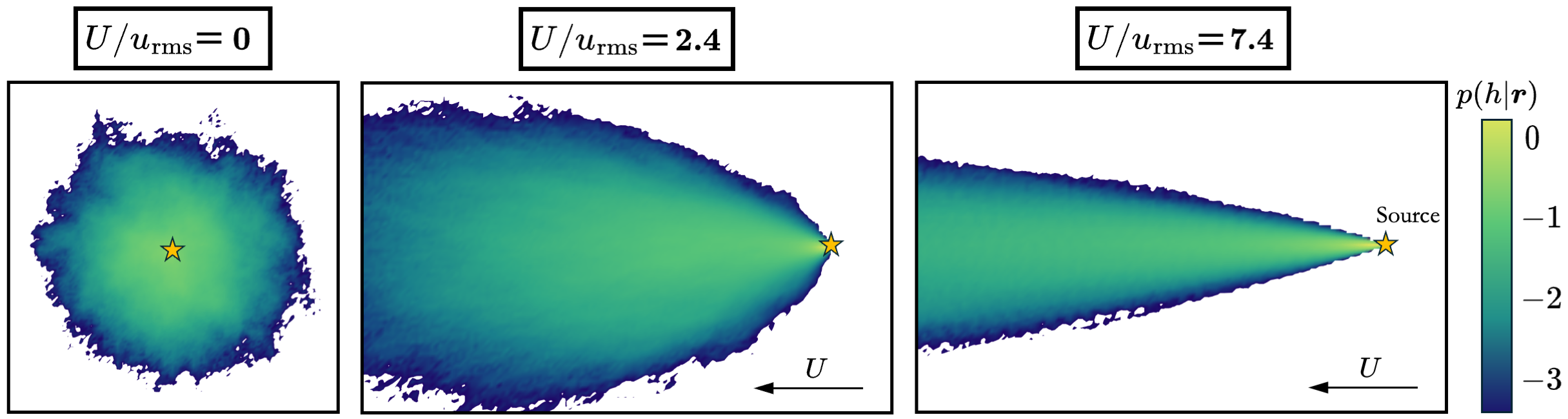}
\caption{\justifying \textbf{Empirical likelihoods obtained from the DNS data.} Probability maps of making a detection in the 2D scenarios considered in the main text. They differ in terms of the mean wind intensity, from left to right: $U/u_{\rm rms}=0$, $U/u_{\rm rms}=2.4$, and $U/u_{\rm rms}=7.4$. These are the models of the environment $p$ the agents use to interpret their observations, also known as \emph{empirical likelihood} in Bayesian jargon. They result from the coarse-graining of the odors' trajectories obtained from the DNS, which have then been thresholded ($c_{\rm thr}=10$), and finally averaged over time.}
\label{fig:empLk_cThr10}
\end{figure*}

\begin{figure*}[ht!]
\centering
\includegraphics[width=\textwidth]{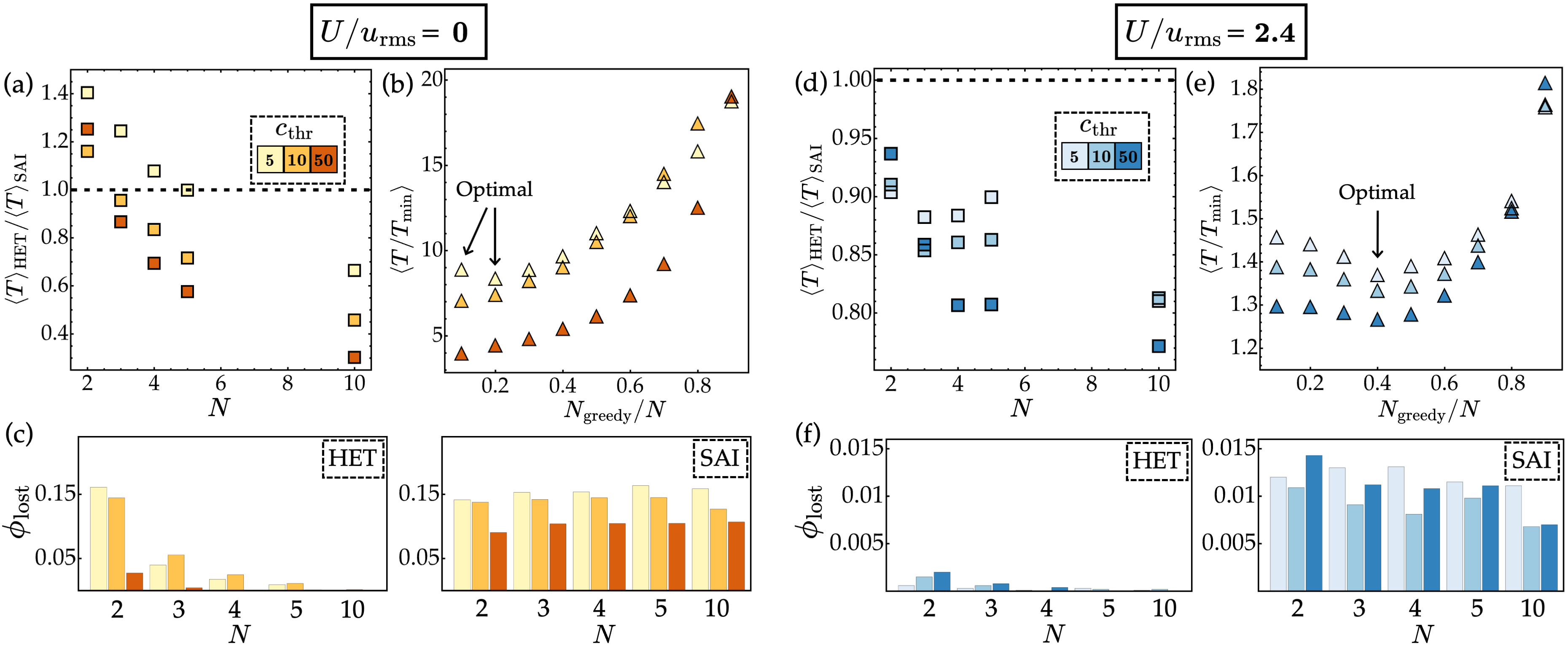}
\caption{\justifying \textbf{Results at varying agents' odor sensitivity.} Results obtained changing the threshold $c_{\rm thr}$ on the number of odor particles detectable by the agents in the 2D setup when in absence of a mean wind [panels (a)-(c)] or in presence of it [$U/u_{\rm rms}=2.4$, panels (d)-(f)].
(a),(d) Ratio between the mean arrival time of the first agent at the source using the optimal fraction of greedy agents ---as defined, in the case of ten agents, by the minima in panels (b) and (e)--- and the one obtained with a SAI swarm, as a function of the number of agents $N$. The arrival time statistics shown here are conditioned on successful episodes, while panels (c) and (f) show the fraction of episodes $\phi_{\rm lost}$ in which the search fails ($T>T_{\rm max}=1400$). Colors denote three different agents' detection thresholds $c_{\rm thr}=\{5,10,50\}$. Note that the data reported here for $c_{\rm thr}=10$ correspond to the ones shown in Fig.~4 of the main text. This comparison shows that the results qualitatively do not depend on the agents' detection threshold, and policy heterogeneity is always beneficial in reaching the odor source faster. Data presented here have been obtained by averaging over $10^4$ episodes.}
\label{fig:vary_cThr}
\end{figure*}

\begin{figure*}[ht!]
\centering
\includegraphics[width=0.3\textwidth]{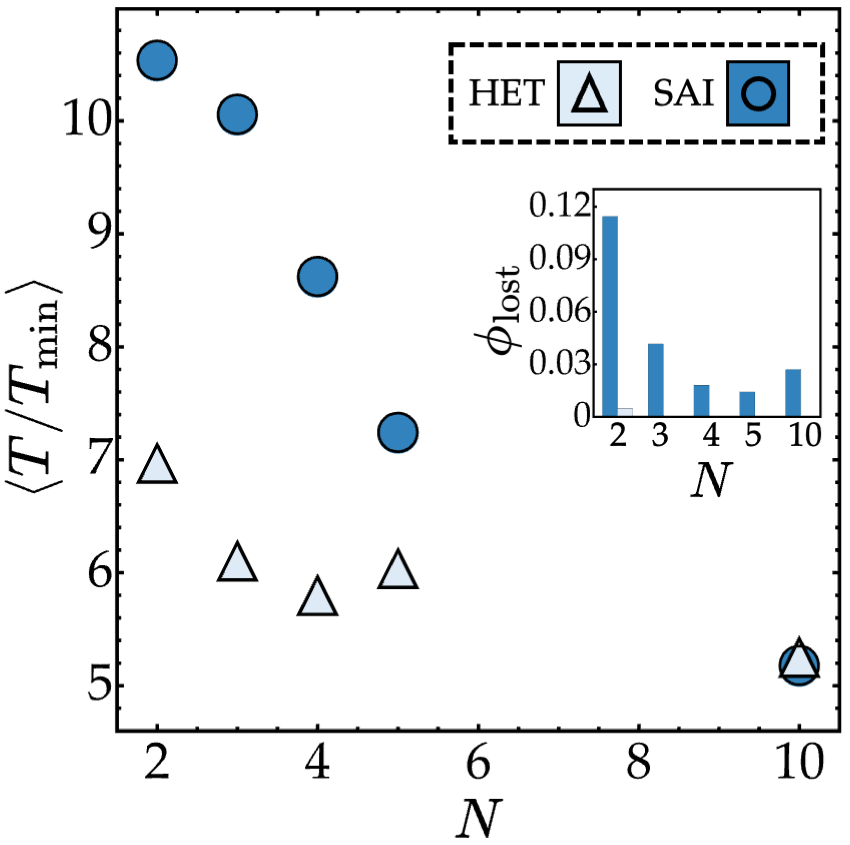}
\caption{\justifying \textbf{Heterogeneity benefits the whole group.} Let us assume that once the first agent finds the source, the shared belief collapses to a delta function in the source position, such that all the other agents can switch to greedy behavior and reach it in a time equal to their Manhattan distance. This allows us to provide a new measure of performance that accounts for the whole group of agents and not only the individual's success. To this end, the results obtained in the same 3-D setup presented in the main text are reported in this figure. Here, we show the mean arrival time of the last agent as a function of the number of agents $N$ composing the SAI swarm (dark blue circles) or the optimal HET configuration (light blue triangles). The arrival time statistics shown here are conditioned on successful episodes, while the inset shows the fraction of episodes $\phi_{\rm lost}$ in which the search fails ($T>T_{\rm max}=2\cdot10^3$). Note that only SAI is visible since for HET swarms $\phi_{\rm lost}<0.01$. 
Remarkably, the whole group benefits from a division of labor as the last agent in HET swarms systematically manages to arrive earlier than the last one among purely SAI agents, especially when the group comprises just a few agents ($N\leq5$).
Data presented here have been obtained by averaging over $5250$ episodes.}
\label{fig:meanTime_last}
\end{figure*}

\end{document}